\documentclass[lettersize,journal]{IEEEtran}
\usepackage{amsmath,amsfonts}
\usepackage{algorithmic}
\usepackage{algorithm}
\usepackage{array}
\usepackage[caption=false,font=normalsize,labelfont=sf,textfont=sf]{subfig}
\usepackage{textcomp}
\usepackage{stfloats}
\usepackage{url}
\usepackage{verbatim}
\usepackage{graphicx}
\usepackage{cite}
\usepackage{pifont}
\usepackage{textcomp,mathcomp}
\usepackage{amsmath,amssymb,amsfonts}
\usepackage{upgreek}
\usepackage{multirow}
\usepackage[colorlinks=true,linkcolor=blue,urlcolor=blue,citecolor=blue]{hyperref}
\usepackage[colorinlistoftodos]{todonotes}

\begin{document}

\title{
Metering Error Estimation of Fast-Charging Stations Using Charging Data Analytics
}

\author{Kang Ma, Xiulan Liu, Xi Chen, Xiaohu Liu, Wei Zhao, Lisha Peng, \textit{Senior Member, IEEE},\\ Songling Huang, \textit{Senior Member, IEEE}, Shisong Li, \textit{Senior Member, IEEE}
\thanks{Kang Ma, Xiaohu Liu, Wei Zhao, Lisha Peng, Songling Huang, and Shisong Li are with the Department of Electrical Engineering, Tsinghua University, Beijing 100084, China. Xiulan Liu and Xi Chen are with the Electric Power Research Institute, State Grid Beijing Electric Power Company, Beijing 100045. }
\thanks{Email: shisongli@tsinghua.edu.cn}
}
%

\markboth{}{}


\maketitle

\begin{abstract}
Accurate electric energy metering (EEM) of fast charging stations (FCSs), serving as critical infrastructure in the electric vehicle (EV) industry and as significant carriers of vehicle-to-grid (V2G) technology, is the cornerstone for ensuring fair electric energy transactions. Traditional on-site verification methods, constrained by their high costs and low efficiency, struggle to keep pace with the rapid global expansion of FCSs. In response, this paper adopts a data-driven approach and proposes the measuring performance comparison (MPC) method. By utilizing the estimation value of state-of-charge (SOC) as a medium, MPC establishes comparison chains of EEM performance of multiple FCSs. Therefore, the estimation of EEM errors for FCSs with high efficiency is enabled. Moreover, this paper summarizes the interfering factors of estimation results and establishes corresponding error models and uncertainty models. Also, a method for discriminating whether there are EEM performance defects in FCSs is proposed. Finally, the feasibility of MPC method is validated, with results indicating that for FCSs with an accuracy grade of 2\%, the discriminative accuracy exceeds 95\%. The MPC provides a viable approach for the online monitoring of EEM performance for FCSs, laying a foundation for a fair and just electricity trading market.
\end{abstract}


% ...
\section{Introduction}\label{Section: introduction}
\begin{figure}[tp!]
    \centering
    \includegraphics[width=0.5\textwidth]{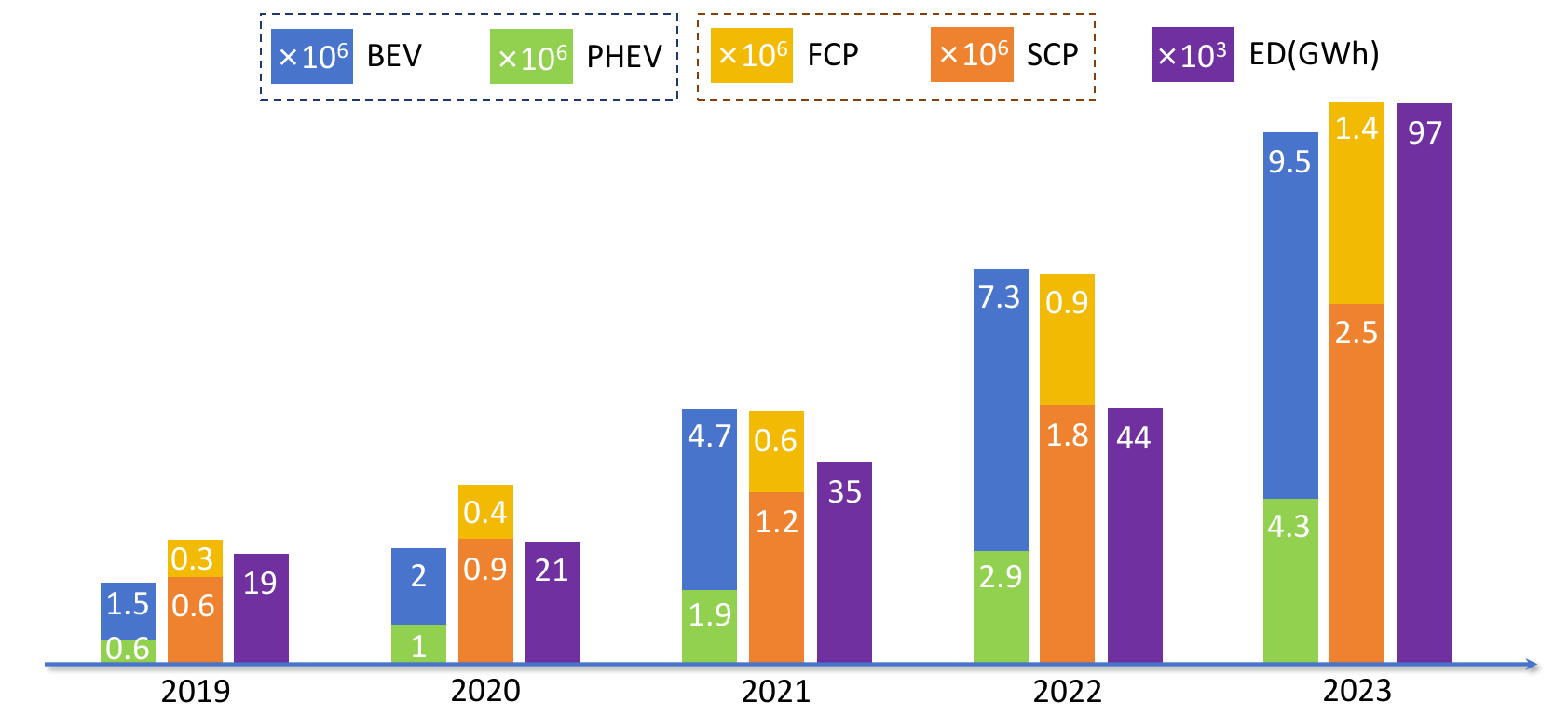}
    \caption{Worldwide statistical data of EV industry, including the sales number of battery electric vehicles (BEV) and plug-in hybrid electric vehicles (PHEV), the number of publicly available fast charging points (FCP) and slow charging points (SCP), and the electricity demand (ED, GWh) by the end of 2023.}
    \label{fig: statistical data of EV industry}
\end{figure}

\IEEEPARstart{A}{s} an environmentally-friendly means of transportation, the electric vehicle (EV) is crucial for addressing the prominent issues faced by humanity, that is, the shortage of fossil fuels and climate change~\cite{review-of-EV-2-202108,review-of-EV-1-202209}. As shown in Fig.~\ref{fig: statistical data of EV industry}, the EV industry has achieved rapid development globally. In order to alleviate the ‘range anxiety’ of EV users and promote the further development of the EV industry, governments around the world have accelerated the construction of infrastructure such as charging stations~\cite{charging-station-2-202405,charging-station-1-202410}. Among them, DC charging stations (also known as fast charging stations, FCSs) have been favored by EV users due to the high charging power. Moreover, as an important carrier of vehicle-to-grid (V2G) technology, the FCS is an indispensable device in future distributed grid~\cite{V2G-1-202204,V2G-2-202301,V2G-3-202407}.
According to the statistical data of the International Energy Agency (IEA), by the end of 2023, the number of publicly available fast charging points worldwide has reached 1.4 million. Including the United States, China, and the European Union, major EV-using countries have almost established a complete fast charging service system. Behind the large-scale FCSs is a huge electricity trading market. The accuracy of EEM at FCS is related to the immediate interest of power grid companies and EV users. Especially for power grid companies, inaccurate EEM will cause massive energy waste and economic losses. Therefore, the accurate EEM value of FCSs is the only way to avoid disputes between power grid companies and EV users.

Due to the impact of complex operating environments and components aging, the EEM error of FCSs is inevitable. Therefore, it is necessary to regularly verify the EEM accuracy of FCSs to guide the calibration work. In China, FCSs have been included in the metering instruments subject to mandatory verification (verification period not exceeding 3 years). As a consequence, domestic research on EEM error verification technology of FCSs started earlier, with the development of field calibrator of EEM error beginning in the 2010s. According to the load type, the field verification is divided into real-load and virtual-load verification~\cite{field-calibration-1-201911}. The real-load or virtual-load field calibrator with high accuracy of EEM is used to meter the same load as the FCS being verified. By comparing the EEM values, the EEM error of the FCS is calculated by
\begin{equation}
    \gamma = \frac{{{E_1} - {E_0}}}{{{E_0}}} \times 100\%,
\end{equation}
where $E_1$ is the EEM value of the FCS, and $E_0$ is the EEM value of the field calibrator. Compared to the real-load verification, the virtual-load verification divides the voltage and current into different circuits, thereby not generating any electric energy loss. Therefore, from the perspective of energy conservation, virtual-load verification has advantages. Domestic scholars have focused their research on the development of field calibrators, aiming to improve hardware and software performance to enhance their intelligence, portability, and anti-interference capabilities. Overseas, in light of the rapid increase in the number of EVs and FCSs, the National Institute of Standards and Technology (NIST) initiated a research project in March 2023 on the development of field calibrator for EEM accuracy of FCSs. Meanwhile, many countries have successively started research on laboratory verification techniques for DC energy meters, and are actively promoting the formulation of relevant laws and regulations~\cite{abroad-reserach-1-202201}.


To ensure the accuracy of field verification results, the EEM value of the field calibrator should not be too small, which leads to disadvantages of high time cost and low efficiency. According to on-site experience, the verification duration for an FCS is approximately 30 minutes. Therefore, the field verification method cannot meet the verification needs of massive FCSs. In view of that, the scholars are also actively exploring remote verification methods. The existing technical proposal involves using a multi-channel switching device to remotely switch the connection of the field calibrator and a cluster of FCSs, thereby achieving remote verification of EEM accuracy~\cite{remote-calibration-1-201909}. However, like the field verification method, this method essentially belongs to a ‘point-to-point’ approach, whose inherent drawbacks of high time cost and low efficiency remain unavoidable.

It is worth noting that the dilemmas faced in EEM error verification of FCSs also exist in accuracy verification of electric meters~\cite{energymeter-cal-1-201507,energymeter-cal-0-20200312,energymeter-cal-2-202401}, voltage transformers~\cite{vol-cal-1-201710}, current transformers~\cite{cur-cal-1-201706,cur-cal-2-201711}, and other meters in the power system. In these fields, scholars have introduced data-driven methods to monitor the accuracy of field meters, thereby saving the verification costs. Specifically, within the topological framework of the power grid, electrical topology connections have been established between different meters, and the theoretical model of the relationships between the measurement values is established. Based on this theoretical model, combined with big data of measurement, the measurement accuracy can be evaluated. Here, the question is can we establish the electrical topology relationship of different FCSs? Based on the measurement data of EVs and FCSs, this paper establishes comparison chains for the EEM performance between multiple FCSs by using the state-of-charge (SOC) value estimated by EVs' battery management system (BMS) as an intermediary. Lastly, the EEM errors of multiple FCSs are estimated by the method of metering performance comparison (MPC). Specifically, assuming that the capacity of the EV's battery pack is stable and the SOC estimation is accurate. For every 1\% increase in SOC, the corresponding charging energy value remains constant. Here, this value is defined as battery pack energy density (BPED), i.e.
\begin{equation}\label{Eq: BPED-S1}
    {E_{\mathrm{d}}} = \frac{E}{{\Delta {\mathrm{SOC}}}},
\end{equation}
where $E$ is the charging energy, $\Delta {\mathrm{SOC}}$ is the change value of SOC, with the scale of \%. Hence, when an EV is charged at FCS-A, the EEM value of FCS-A is $E_{\mathrm{A}}$, and the change value of SOC is $\Delta {\mathrm{SOC}_{\mathrm{A}}}$. Subsequently, when the same EV is charged at FCS-B, the EEM value of FCS-B is $E_{\mathrm{B}}$, and the change value of SOC is $\Delta {\mathrm{SOC}_{\mathrm{B}}}$. The relative EEM error of FCS-B to FCS-A is given by
\begin{eqnarray}\label{Eq:relative error-S1}
    {\gamma _{{\mathrm{B}} \to {\mathrm{A}}}} &=& \frac{{{E_{{\mathrm{dB}}}} - {E_{{\mathrm{dA}}}}}}{{{E_{{\mathrm{dA}}}}}} \nonumber\\
    &=& \frac{{\frac{{{E_{\mathrm{B}}}}}{{\Delta {\mathrm{SO}}{{\mathrm{C}}_{\mathrm{B}}}}} - \frac{{{E_{\mathrm{A}}}}}{{\Delta {\mathrm{SO}}{{\mathrm{C}}_{\mathrm{A}}}}}}}{{\frac{{{E_{\mathrm{A}}}}}{{\Delta {\mathrm{SO}}{{\mathrm{C}}_{\mathrm{A}}}}}}} \times 100\% . 
\end{eqnarray}
So far, the electrical topology between multiple FCSs has been established. Combined with the big data of EVs charging, the EEM errors of multiple FCSs can be estimated.

The remaining sections of this paper are organized as follows: Section \ref{Section: Principle} introduces the principles of the MPC method, Section \ref{Section: Interfering factors} investigates the interfering factors affecting the accuracy of the estimation results, Section \ref{Section: algorithm and uncertainty} provides the estimation algorithm and analyzes the uncertainty of the estimation results, Section \ref{Section: validation} conducts the validation of the estimation model, and Section \ref{Section: conclusion} presents the conclusions of this paper.

\section{Principle of the proposed data-driven metering error estimation method}\label{Section: Principle}
\begin{figure}[tp!]
    \centering
    \includegraphics[width=0.5\textwidth]{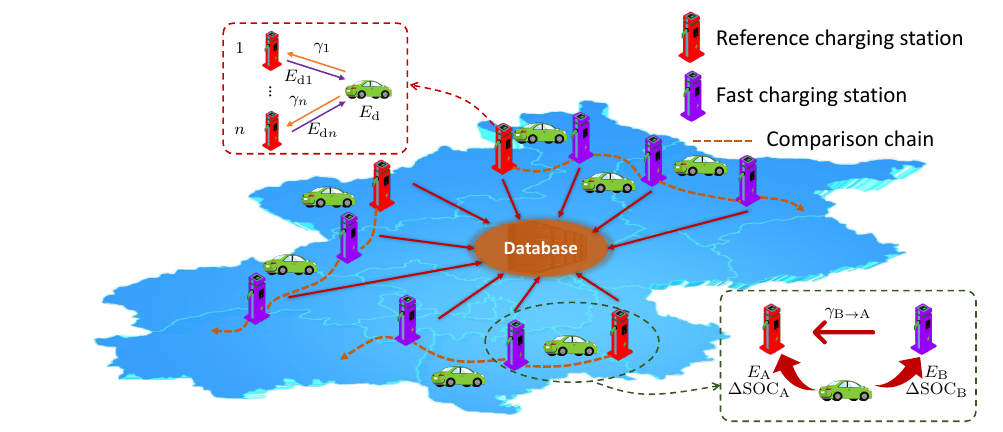}
    \caption{Principle of the estimation method for FCSs' EEM accuracy based on charging data.}
    \label{fig: principle of the method}
\end{figure}

\begin{figure}[tp!]
    \centering
    \includegraphics[width=0.425\textwidth]{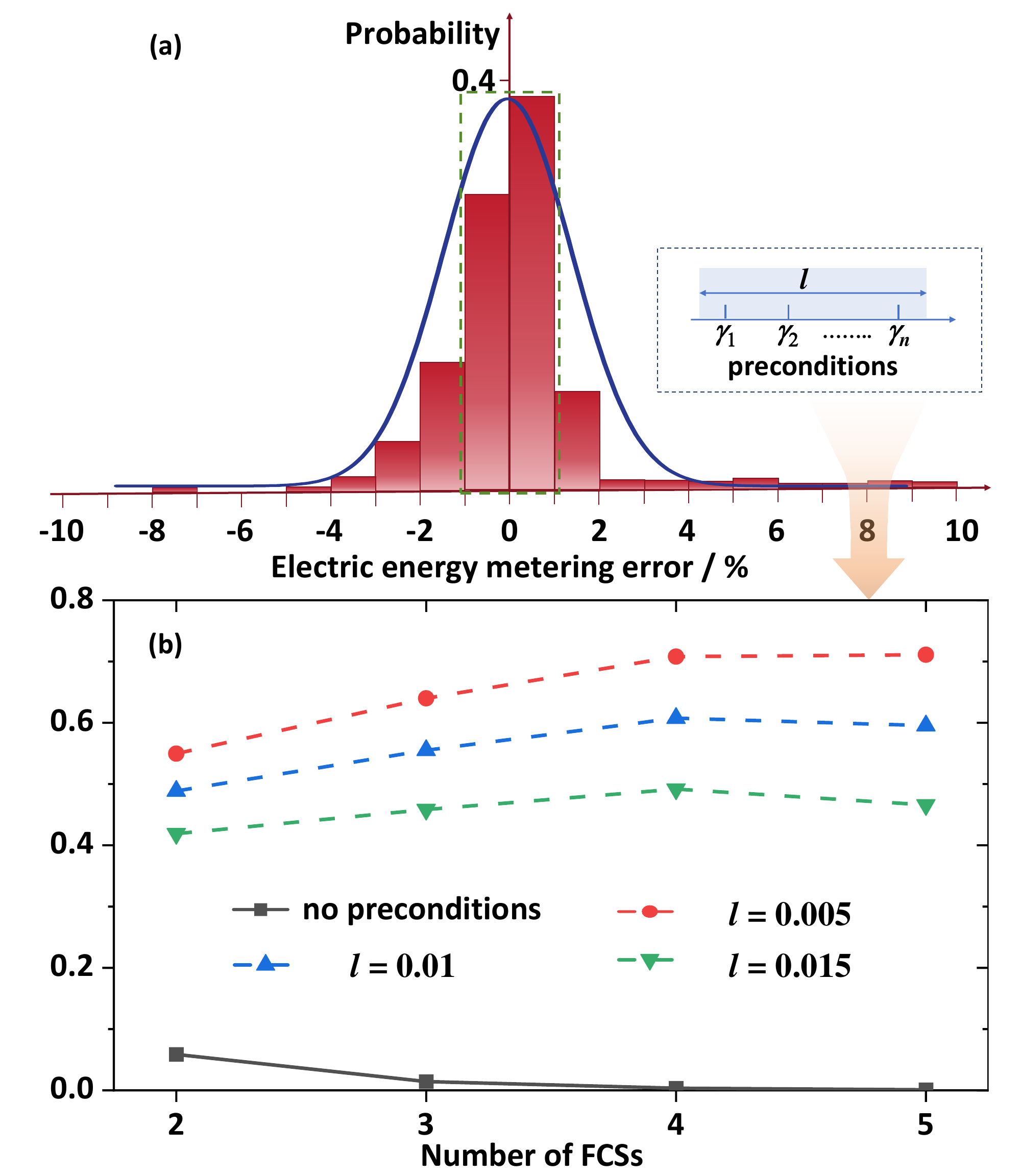}
    \caption{Determination of the RCSs. (a) is the normal distribution of EEM error of on-site FCSs. The number of FCSs is 3500 produced by 10 manufacturers. According to the fitting result, the average value of this normal distribution is approximately 0, and the standard deviation equals 1.62\%. (b) is the probability of EEM errors in FCSs cluster. The horizontal coordinate represents the number of FCSs in the cluster, and the vertical coordinate represents the probability that the EEM errors of all FCSs are within the range of [-0.01,0.01]. Different curves represent the probability under different preconditions.}
    \label{fig: distribution of error}
\end{figure}

As mentioned in Section \ref{Section: introduction}, the principle of the MPC method for EEM errors of FCSs is shown in Fig.~\ref{fig: principle of the method}. Under the premise that the capacity of the EV's battery pack is stable and the SOC estimation is accurate, according to big data of EVs charging records, reference charging stations (RCSs) are selected, and the EEM error of RCSs is estimated first. Then, multiple comparison chains are built with the beginning of RCSs. For Equation (\ref{Eq:relative error-S1}), assuming that the EEM error of FCS-A is $\gamma _{\mathrm{A}}$, the EEM error of FCS-B is given by
\begin{equation}\label{Eq:error cal-S2}
    {\gamma _{\rm{B}}} = {\gamma _{{\rm{B}} \to {\rm{A}}}} + \frac{{{E_{{\rm{dB}}}}}}{{{E_{{\rm{dA}}}}}}{\gamma _{\rm{A}}}. 
\end{equation}
Herein, the EEM errors of all FCSs are calculable.
Presently, the question is how to select the RCSs. As shown in Fig.~\ref{fig: distribution of error}(a), according to the on-site statistical data, the EEM errors of FCSs follow a normal distribution with an average value of 0. For one charging record of an EV charging at a FCS, the relationship between the estimation and true value of BPED is depicted as
\begin{equation}\label{Eq: error-S2}
    {{\hat E}_{{\rm{d}}i}} = \left( {1 + {\gamma _i}} \right){E_{\rm{d}}},
\end{equation}
where ${\hat E}_{{\rm{d}}i}$ is the estimation value, $E_{\mathrm{d}}$ the true value and $\gamma _i$ the EMM error of the FCS-$i$. Hence, when the same EV has charging records at multiple FCSs, the average value of BPED estimation is 
\begin{equation}\label{Eq: RCS-error}
    {{\hat E}_{\rm{d}}} = {E_{\rm{d}}} + \frac{{\sum\limits_{i = 1}^n {{\gamma _i}} }}{n}{E_{\rm{d}}}. 
\end{equation}
$n$ is the number of FCSs. Ideally, as long as $n$ is large enough, the value of ${\sum\limits_{i = 1}^n {{\gamma _i}} }$ will infinitely approach 0, that is the estimation value will infinitely approach the true value. Furthermore, the EEM errors of these FCSs can be fixed according to Equation (\ref{Eq: error-S2}), and these FCSs can be considered as RCSs. However, the reality is that few FCSs serve the same EV. Most likely, it will lead to a significant error in the estimation value of BPED. To address this issue, we need to add some screening criteria for FCSs to improve the estimation accuracy of the BPED.

Actually, it is expected to select a cluster of FCSs with EEM errors close to 0. By utilizing the charging records of an EV at this cluster of FCSs, a more accurate estimation of BPED can be obtained. According to Fig.~\ref{fig: distribution of error}(b), if no preconditions are added, the probability that the EEM errors of $n$-FCS cluster are all within the range of [$-\gamma _0$,$\gamma _0$] is quite small. Therefore, the precondition that the relative EEM errors between any two FCSs in the cluster is smaller than a specific value $l$ is added to improve such probability. As shown in Fig.~\ref{fig: distribution of error}(b), such probability is significantly improved under this precondition, especially when $l$ is small. In practice, by selecting the appropriate value of $l$ and the size of the FCSs cluster, the suitable cluster can be selected and the EEM errors of RCSs can be fixed by the aforementioned method.

So far, the EEM errors of large-scale FCSs can be estimated by utilizing the big data of charging records, ideally. But in reality, the capacity of EV's battery pack is time-variant. Meanwhile, the accurate estimation of SOC is a very challenging task due to the high nonlinearity of EV's battery system. Hence, the data pre-processing is vital to accurate estimation. Afterward, the interfering factors of estimation accuracy are analyzed, and the method of data pre-processing is introduced.

\section{Interfering factors of estimation accuracy}\label{Section: Interfering factors}

\subsection{Repeatability error of BPED}\label{Subsection: Repeatability error of BPED}
\begin{figure}[tp!]
    \centering
    \includegraphics[width=0.45\textwidth]{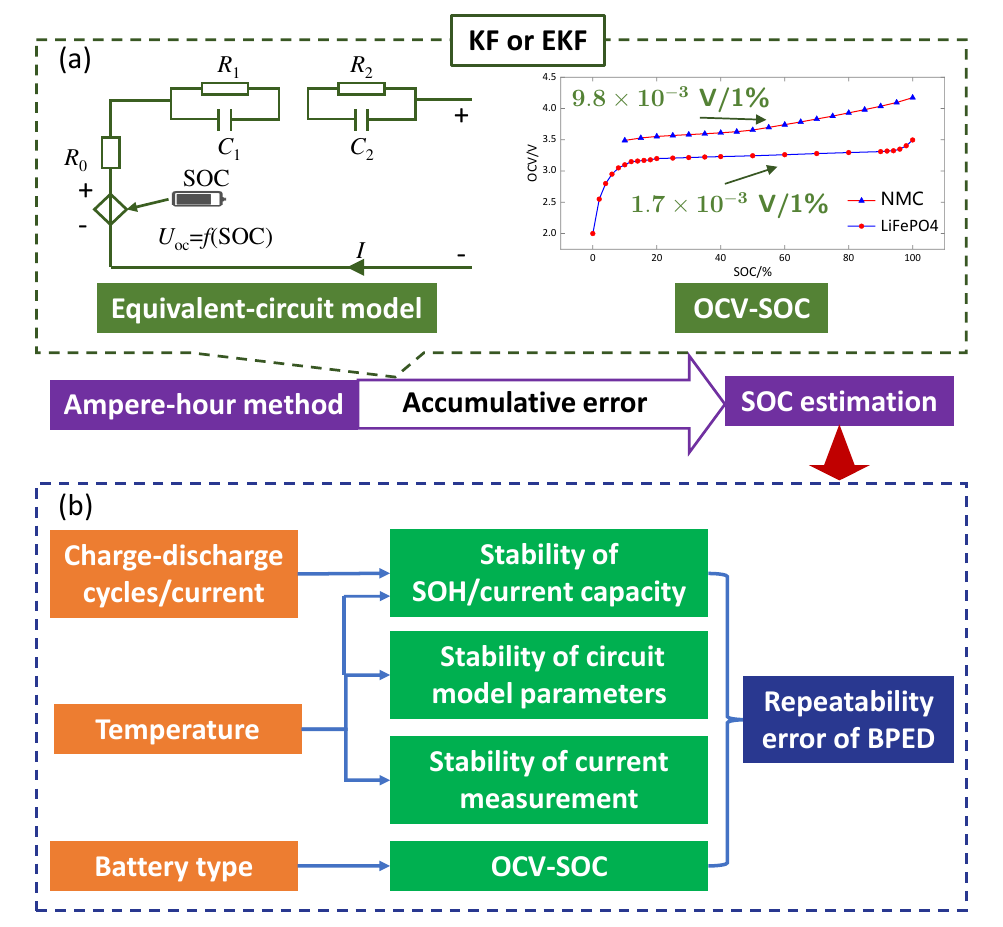}
    \caption{(a) is the method based on KF or EKF. In which, the state equations are obtained according to the equivalent-circuit model and OCV-SOC curve~\cite{OCV-SOC-202401}. The accumulative error of the ampere-hour method is compensated. (b) is the interfering factor on the repeatability error of BPED, including the stability of state-of-healthy (SOH)/current capacity, circuit model parameters, current measurement, and the characteristics of OCV-SOC curve. Moreover, the stability of SOH/current capacity is primarily influenced by the number of charge-discharge cycles and temperature. The stability of circuit model parameters is primarily affected by temperature variations. The stability of current measurement is dependent on the stability of the onboard current sensor, which is mainly influenced by temperature drift. The OCV-SOC curve is determined by the intrinsic characteristics of the battery. Generally speaking, the OCV-SOC curve characteristics of different battery types exhibit significant deficiencies.}
    \label{fig: interfering factors of BPED}
\end{figure}

In the MPC method, the constancy of BPED serves as a prerequisite for establishing accurate comparison chains. According to Equation (\ref{Eq: BPED-S1}), the stability of BPED, which refers to the capacity to maintain equal changes in SOC when the charging/discharging energy is the same, is closely related to the SOC estimation algorithm and status of the battery pack. 
To date, the SOC estimation algorithms currently used in EV products are primarily based on the Kalman filter/extended Kalman filter (KF/EKF)~\cite{Kalman-2-202210,Kalman-1-202301,Kaerman-3-202105}.
As shown in Fig.~\ref{fig: interfering factors of BPED}(a). The KF/EKF method is grounded in the ampere-hour (Ah) method. The equation of the Ah method is given by~\cite{SOC-model-201910}

\begin{equation}
    {\rm{SOC}}\left( t \right) = {\rm{SOC}}\left( {{t_0}} \right) + \frac{1}{{{C_{\rm{r}}} \times {\rm{SOH}}}}\int_{{t_0}}^{{t_0} + t} {I{\rm{d}}} \left( \tau  \right),
\end{equation}
where ${\rm{SOC}}\left(t_0\right)$ is the SOC value at at initial state. $C_{\rm{r}}$ is the rated capacity. SOH is the state-of-healthy of the battery pack, which is defined as the ratio of the current capacity to the rated capacity. $I$ is the charging current. Owing to the absence of calibration and feedback mechanisms in the Ah method, cumulative errors progressively accumulate over time, leading to a significant deviation of the estimated value from the true value. To mitigate the cumulative errors, the KF or EKF is employed to rectify the estimation result of the Ah method.
According to Fig.~\ref{fig: interfering factors of BPED} (b), the repeatability error of BPED is affected by the number of charge-discharge cycles, temperature, and battery type. Specifically,

\textcircled{1} Charge-discharge cycles/current:

The variation of lithium-ion battery (the most common battery type for EVs) capacity with the number of charge-discharge cycles is studied in \cite{cycletoSOH}. The capacity of the onboard battery pack gradually decreases with the increase of charge-discharge cycles. Nonetheless, when the cycle number is less than 500, the decay in battery capacity is less than 0.2~\% per every 100 cycles. Therefore, in order to mitigate or avoid the impact of charge-discharge cycles, an upper limit should be set for the time interval of charging data used in the estimation model.


The impact of charging current on the capacity is studied in \cite{tem-to-capa-1-201302} and \cite{tem-to-capa-2-201610}. Usually, the capacity will decrease with the increase of charging current. Especially, in low (below 20~\textcelsius) and high (above 40~\textcelsius) temperature environments, the impact of charging current is more severe.

In addition, the charging current will affect the conversion efficiency. The conversion efficiency is defined as the ratio of the actual charging capacity to the EEM value of the FCS. Due to the existence of line losses in the cable of the charging gun, the conversion efficiency is typically less than 1. By equating the cable to a resistor, the conversion efficiency is given by
\begin{equation}\label{Eq: eta-S3}
    \eta  = \left( {1 - \frac{{IR}}{U}} \right) \times 100\% . 
\end{equation}
Where $U$ represents the charging voltage, $I$ represents the charging current, and $R$ represents the resistance value of the cable. Subsequently, when evaluating the uncertainty, it is necessary to take into account the influence of conversion efficiency.

\textcircled{2} Temperature:

According to Fig.~\ref{fig: interfering factors of BPED} (b), the influence of temperature includes:
\begin{enumerate}
    \item SOH/current capacity

    Existing research has demonstrated that ambient temperature is a significant factor affecting the capacity of on-board battery packs~\cite{tem-to-capa-1-201302,tem-to-capa-2-201610,hum-to-capa-1-202406,hum-to-capa-2-202406}. Referring to \cite{tem-to-capa-1-201302} and \cite{tem-to-capa-2-201610}, when the temperature is below 20~\textcelsius, the capacity reduction of lithium-ion batteries is more severe. While the temperature exceeds 40~\textcelsius, the capacity exhibits a slight decline.

    
    \item Circuit model parameters

    Currently, battery models encompass electrochemical models and equivalent-circuit models. Considering the trade-offs between simulation accuracy, complexity, and parameter identification difficulty, equivalent-circuit models have become commonly used battery models in SOC estimation algorithms, particularly the 2-order Thevenin model. As illustrated in Fig.~\ref{fig: interfering factors of BPED} (a), The parameters of the 2-order Thevenin model include: the ohmic internal resistance ($R_0$), the polarization resistances ($R_1$ and $R_2$), and the polarization capacitors ($C_1$ and $C_2$). According to reference~\cite{tem-to-parameters-202101}, a decrease in temperature leads to a significant increase in the ohmic resistance value. The value of $R_0$ at -10~\textcelsius  is twice of that at 25~\textcelsius. Moreover, with variations in temperature, both the polarization resistances and the polarization capacitance values undergo substantial fluctuations.
    
    \item Current measurement

   The accuracy of the Ah method is highly dependent on the measurement accuracy of the sensors within the battery pack, particularly the current sensors. Considering the space utilization and reliability requirements of EVs, onboard battery packs typically require current sensors that are compact, cost-effective and have strong anti-interference capabilities. Commonly used sensors include Hall-effect current sensors and shunt resistors. These sensors are designed with temperature considerations and incorporate corresponding compensation measures, allowing their temperature coefficients to be controlled within 0.1\%/\textcelsius~\cite{tem-to-sensor1-202004}.
\end{enumerate}

\textcircled{3} Battery type:

Lithium-ion batteries employed in EVs encompass various types, including LiNiMnCoO2 (NMC) batteries, NCA batteries, LiFePO4 batteries, and etc~\cite{Libattery-overview1-202112,Libattery-overview2-202306}. These batteries exhibit differences in energy density, cycling performance, and safety characteristics. It's noted that the characteristics of OCV-SOC curve for different types is different. The OCV-SOC curves for NMC batteries and LiFePO4 batteries are shown in Fig.~\ref{fig: interfering factors of BPED} (a). Due to the flat region in the middle segment of the OCV-SOC curve for LiFePO4 batteries, the prediction of SOC by OCV method will result in significant errors. Consequently, in the MPC method, to ensure the accuracy of comparison chains, the charging data from LiFePO4 batteries is not recommended.


In summary, the stability of BPED is subject to numerous interference factors, and the mechanisms are complex, making it difficult to establish a theoretical model for repeatability errors. Consequently, the repeatability error of BPED is quantified through experimental methods. Here, two EVs from different manufacturers are selected as experimental subjects and multiple charging tests are conducted on the two EVs. The distribution characteristics of BPED values for the two test EVs are statistically analyzed, separately.
\begin{figure}[tp!]
    \centering
    \includegraphics[width=0.45\textwidth]{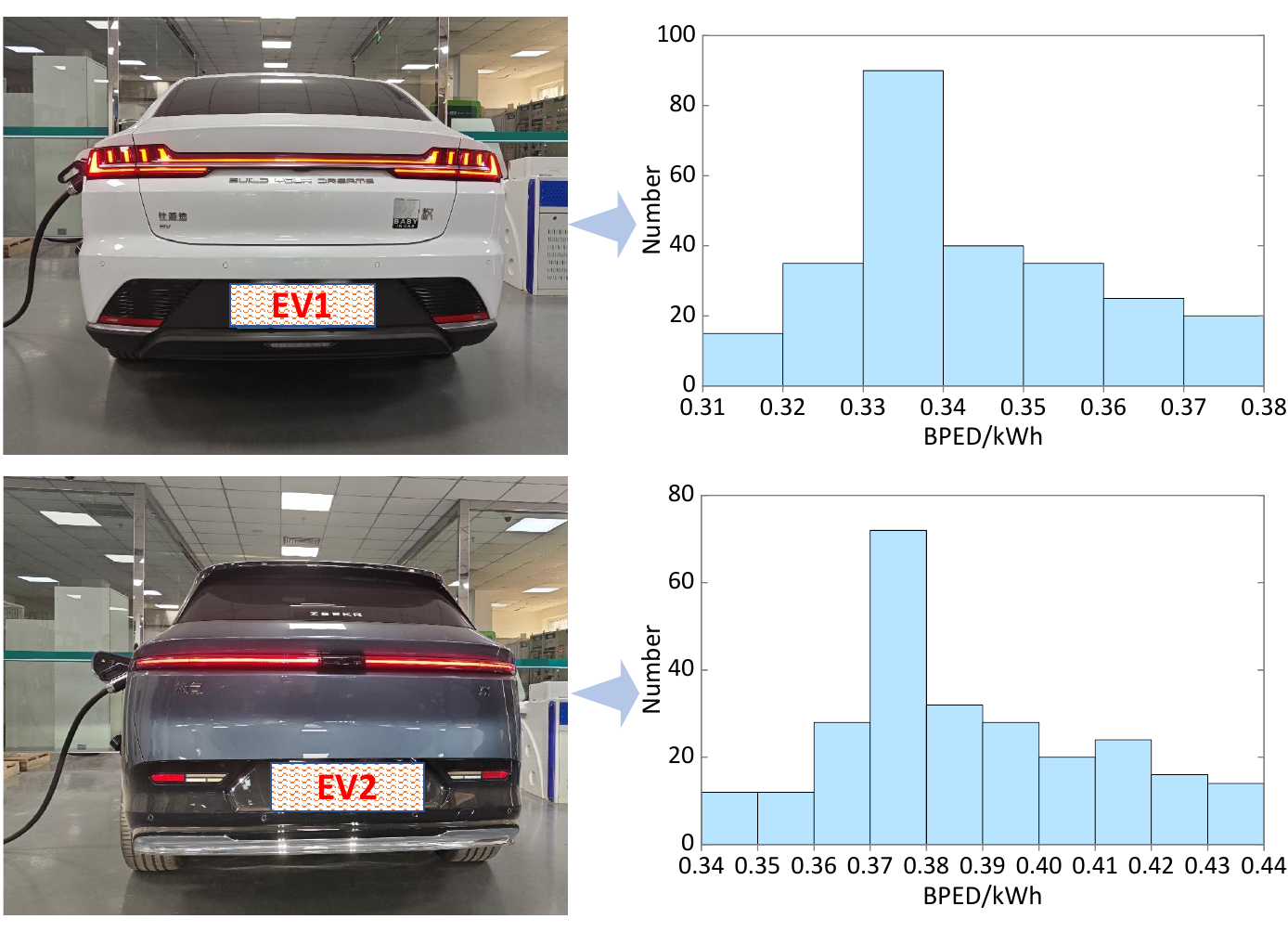}
    \caption{Experimental results. The charging mode of the test FCS is constant current. The resistance value of the charging gun is 3\,${\mathrm{m}}\Omega$. Before the charging test, the SOC of test EVs is controlled under 10\%. For each test Ev, the number of charging-discharging cycles is 3.}
    \label{fig: experiment result}
\end{figure}

As shown in Fig.~\ref{fig: experiment result}, the BPED values of the test EVs exhibit the normal distribution. For EV1, within the statistical interval, the range of charging current varies from 42.8 to 44.5~A, the average temperature of the battery pack is 28.5~\textcelsius. According to the fitting results, the mean value of BPED is 0.34\,kWh, and the standard deviation is 0.017\,kWh. For EV2, the charging current ranges from 39.7 to 43.4\,A, the average temperature of the battery pack is 29.4~\textcelsius. The mean value of BPED is 0.38\,kWh, and the standard deviation is 0.022\,kWh. That is, the relative repeatability error of BPED is less than 6\%.


\subsection{Quantization error of SOC}\label{Subsection: SOC quantization}

\begin{figure}[tp!]
    \centering
    \includegraphics[width=0.5\textwidth]{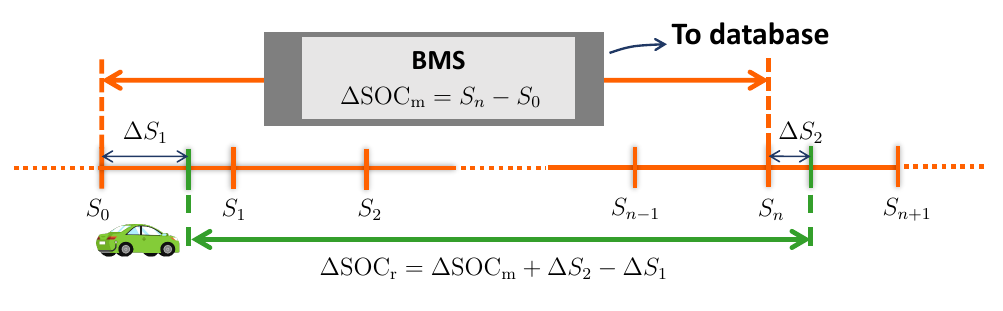}
    \caption{Schematic diagram of SOC quantification error. The green line represents the real change of SOC, whereas the orange line denotes the change of SOC as calculated by the BMS.}
    \label{fig: quantification error}
\end{figure}

In reality, the sampling frequency of operational FCSs doesn't exceed 1/60~Hz, and the resolution of SOC in the onboard BMS is only 1\%, which leads to quantization error of SOC. As illustrated in Fig.~\ref{fig: quantification error}, the D-value between $\Delta {\mathrm{SOC_r}}$ and $\Delta {\mathrm{SOC_m}}$ is referred to the SOC quantization error. Assuming the charging energy is $E$, the estimation error of the BPED due to the SOC quantization error is given by
\begin{equation}\label{Eq: quantization error-S3}
    e = \frac{{{E_{{\rm{dm}}}} - {E_{{\rm{dr}}}}}}{{{E_{{\rm{dr}}}}}} = \frac{{\Delta {S_2} - \Delta {S_1}}}{{{S_n} - {S_0}}} \times 100\% 
\end{equation}
Typically, the change in SOC is less than 90\% during the EV charging. The value of $\Delta S_2-\Delta S_1$ is a random variable distributed within the interval $\left(-1,1\right)$. Consequently, the error in BPED estimation caused by the SOC quantization error is higher than 2\%, which can't be ignored.
In light of this, a probability model for the SOC quantification error is established. Assuming that $\Delta S_1$ and $\Delta S_2$ follow a uniform distribution over an interval $\left[0,1\right)$, the probability density function is given by
\begin{equation}
    {f_{\Delta S}}(x) = \left\{ {\begin{array}{*{20}{l}}
{1,0 \le x < 1}\\
{0,{\rm{others}}}
\end{array}} \right..
\end{equation}
Meanwhile, $\Delta S_1$ and $\Delta S_2$ are are mutually independent. Thus, the probability distribution for the quantization error $y=\Delta S_2-\Delta S_1$ is given by
\begin{align}\label{Eq: distribution_y-S3}
    {f_{\rm{Y}}}(y) &= \int_{ - \infty }^{ + \infty } {{f_{\Delta {S_1}}}(x){f_{\Delta {S_2}}}(y + x){\rm{d}}x}
    \notag
    \\& = \left\{ {\begin{array}{*{20}{l}}
{1 + y, - 1 < y < 0}\\
{1 - y,0 \le y < 1}\\
{0,{\rm{others}}}
\end{array}} \right..
\end{align}

Furthermore, assuming that the data points uploaded by the FCS are represented as $\left[ {\left( {{S_0},{E_0}} \right),\left( {{S_1},{E_1}} \right), \cdots ,\left( {{S_n},{E_n}} \right)} \right]$, the true value of BPED should satisfy the inequalities:

\begin{equation}
    \left\{ {\begin{array}{*{20}{c}}
{\frac{{{E_1} - {E_0}}}{{{S_1} - {S_0} + 1}} \le {E_{\rm{d}}} \le \frac{{{E_1} - {E_0}}}{{{S_1} - {S_0} - 1}}}\\[6pt]
{\frac{{{E_2} - {E_0}}}{{{S_2} - {S_0} + 1}} \le {E_{\rm{d}}} \le \frac{{{E_2} - {E_0}}}{{{S_2} - {S_0} - 1}}}\\[6pt]
 \cdots \\[6pt]
{\frac{{{E_n} - {E_0}}}{{{S_n} - {S_0} + 1}} \le {E_{\rm{d}}} \le \frac{{{E_n} - {E_0}}}{{{S_n} - {S_0} - 1}}}
\end{array}} \right..
\end{equation}

By solving the aforementioned system of inequalities, the range of the true value of BPED is obtained as $\left[ {{E_{{\rm{d}}\min }},{E_{{\rm{d}}\max }}} \right]$. Thereby, the range of the quantization error is scaled to $[\frac{E}{{{E_{{\rm{d}}\max }}}} - ({S_n} - {S_0}),\frac{E}{{{E_{{\rm{d}}\min }}}} - ({S_n} - {S_0})]$. Let ${y_{\min }} = \frac{E}{{{E_{{\rm{d}}\max }}}} - ({S_n} - {S_0})$, ${y_{\max }} = \frac{E}{{{E_{{\rm{d}}\min }}}} - ({S_n} - {S_0})$, under the prior condition, the conditional probability density function of the SOC quantization error ${f_{y|({y_{\min }} \le y \le {y_{\max }})}}(y)$ can be obtained.

Herein, having considered the SOC quantization error, the value of BPED should be given by the unbiased estimation instead of Equation (\ref{Eq: BPED-S1}). The unbiased estimation is given by Equation (\ref{Eq: expected Ed-S3}), which is detailed in Table \ref{Table: Ed detail}.
\begin{equation}\label{Eq: expected Ed-S3}
    {\rm{E}}({E_{\rm{d}}}) = \int_{{y_{\min }}}^{{y_{\max }}} {\frac{E}{{{S_n} - {S_0} + y}}} {f_{y|({y_{\min }} \le y \le {y_{\max }})}}{\rm{d}}y.
\end{equation}

\begin{table*}[tp!]
\caption{Details of Equation (\ref{Eq: expected Ed-S3}) and (\ref{Eq: Ed2-S4})}
\begin{center}
\begin{tabular}{c|c|c}
\hline
Condition & Equation & Ref\\
\hline
\multirow{2}*{$y_{\mathrm{max}}<0$} & ${\rm{E}}({E_{\rm{d}}}) = \frac{{2E}}{a}\left[ {{y_{\max }} - {y_{\min }} + \left( {1 + {y_0}} \right)\ln \frac{{{y_{\max }} + {y_0}}}{{{y_{\min }} + {y_0}}}} \right]$ & (\ref{Eq: expected Ed-S3}-A)\\
\cline{2-3}
~ & ${\rm{E}}(E_{\rm{d}}^2) = \frac{{2{E^2}}}{a}\left[ {\frac{{1 + {y_{\min }}}}{{{y_0} + {y_{\min }}}} - \frac{{1 + {y_{\max }}}}{{{y_0} + {y_{\max }}}} + \ln \frac{{{y_0} + {y_{\max }}}}{{{y_0} + {y_{\min }}}}} \right]$ & (\ref{Eq: Ed2-S4}-A)\\
\hline
\multirow{2}*{$y_{\mathrm{min}}>0$} & ${\rm{E}}({E_{\rm{d}}}) = \frac{{2E}}{b}\left[ {{y_{\min }} - {y_{\max }} + \left( {1 + {y_0}} \right)\ln \frac{{{y_{\max }} + {y_0}}}{{{y_{\min }} + {y_0}}}} \right]$ & (\ref{Eq: expected Ed-S3}-B)\\
\cline{2-3}
~ & ${\rm{E}}(E_{\rm{d}}^2) = \frac{{2{E^2}}}{b}\left[ {\frac{{1 + {y_0}}}{{{y_0} + {y_{\min }}}} - \frac{{1 + {y_0}}}{{{y_0} + {y_{\max }}}} + \ln \frac{{{y_0} + {y_{\min }}}}{{{y_0} + {y_{\max }}}}} \right]$ & (\ref{Eq: Ed2-S4}-B)\\
\hline
\multirow{2}*{$y_{\mathrm{max}}>0$ and $y_{\mathrm{min}}<0$} & ${\rm{E}}({E_{\rm{d}}}) = \frac{{2E}}{c}\left[ {\left( {1 + {y_0}} \right)\ln \frac{{{y_{\max }} + {y_0}}}{{{y_0}}} + \left( {1 - {y_0}} \right)\ln \frac{{{y_0}}}{{{y_{\min }} + {y_0}}} - {y_{\min }} - {y_{\max }}} \right]$ & (\ref{Eq: expected Ed-S3}-C)\\
\cline{2-3}
~ & ${\rm{E}}(E_{\rm{d}}^2) = \frac{{2{E^2}}}{c}\left[ {1 + \frac{{1 + {y_{\min }}}}{{{y_0} + {y_{\min }}}} - \frac{{1 + {y_0}}}{{{y_0} + {y_{\max }}}} + \ln \frac{{{y_0}}}{{{y_0} + {y_{\max }}}} + \ln \frac{{{y_0}}}{{{y_0} + {y_{\min }}}}} \right]$ & (\ref{Eq: Ed2-S4}-C)\\
\hline

\end{tabular}
\label{Table: Ed detail}
\end{center}
\end{table*}

In which, 
\begin{equation}\label{Eq: y0-S3}
    {y_0} = {S_n} - {S_0},
\end{equation}
\begin{equation}\label{Eq: a-S3}
    a = \left( {{y_{\max }} - {y_{\min }}} \right)\left( {2 + {y_{\max }} + {y_{\min }}} \right),
\end{equation}
\begin{equation}\label{Eq: b-S3}
    b = \left( {{y_{\max }} - {y_{\min }}} \right)\left( {2 - {y_{\max }} - {y_{\min }}} \right),
\end{equation}
\begin{equation}\label{Eq: c-S3}
    c = 2\left( {{y_{\max }} - {y_{\min }}} \right) - \left( {y_{\max }^2 + y_{\min }^2} \right). 
\end{equation}

\section{Estimation algorithm and uncertainty analysis}\label{Section: algorithm and uncertainty}
\subsection{Estimation algorithm}
According to the aforementioned content, the EEM errors estimation algorithm for FCSs is as follows:

\textcircled{1} Inputting the raw charging order data: The raw order data encompasses the charging process data for all FCSs, with the time span of charging orders not exceeding two months (Section~\ref{Subsection: Repeatability error of BPED}). The charging process data include at least the following items: time information, EEM value of the FCS, SOC, temperature, charging current, and voltage.

\textcircled{2} Data pre-processing: The pre-processing includes the following steps:

\begin{enumerate}
    \item For the same EV, the charging process data of each charging order is grouped by charging current to form charging data segments, which will be used for the subsequent identification of RCSs and comparison chains. In each charging data segment, the peak-to-peak value of the charging current does not exceed the set threshold, $\Delta I$ (Section \ref{Subsection: Repeatability error of BPED}).
    \item Exclude charging data segments with an average temperature below 20~\textcelsius \,or above 40~\textcelsius \,(Section \ref{Subsection: Repeatability error of BPED}).
    \item Exclude charging data segments with the SOC-change-value less than the set threshold to reduce the impact of SOC quantization error (Equation (\ref{Eq: quantization error-S3})).
    \item Exclude charging data segments related to EVs with poor stability of expected BPED. Herein, an evaluation method for the stability of expected BPED is introduced. To eliminate the impact of FCSs, charging data segments from the same EV and the same FCS are employed for the stability evaluation. Specifically, if the number of charging data segments is $n$. The stability is characterized by the relative repeatability error, i.e.:
    \begin{equation}
        \delta  = \frac{{\sqrt {\frac{{\sum\limits_i {{{\left( {{E_{\rm{d}}}_i - {{\bar E}_{\rm{d}}}} \right)}^2}} }}{{n - 1}}} }}{{{{\bar E}_{\rm{d}}}}}.
    \end{equation}
    $E_{{\mathrm{d}}i}$ represents the expected value of the BPED corresponding to the $i$-th charging data segment, which is calculated by Equation (\ref{Eq: expected Ed-S3}). ${{\bar E}_{\rm{d}}}$ denotes the average value of $E_{{\mathrm{d}}i}$. It is noteworthy that this data processing initiative is of paramount importance. During the operation, some battery cells of an EV may sustain damage, which can significantly impair the performance of the battery pack. Furthermore, some manufacturers have commenced the provision of battery replacement services, resulting in the same EV exhibiting entirely different battery pack performance across different time periods. Consequently, this data processing measure effectively mitigates substantial fluctuations in BPED caused by the aforementioned phenomena.
\end{enumerate}

\textcircled{3} Identification of the RCSs: According to the method mentioned in Section \ref{Section: Principle}, the RCSs are determined and the EEM errors are calculated by Equation (\ref{Eq: error-S2}). In order to maintain the battery pack state as similar as possible, the D-value of the average charging current and temperature between charging data segments must be less than the predefined threshold. 

\textcircled{4} Identification of the comparison chains: The starting point of each comparison chain is the RCS, and no FCS is repeated within each comparison chain. The termination conditions for comparison chains are as follows: a) the length of the comparison chain reaches the maximum allowable value, and b) a RCS (other than started RCS) appears. Moreover, the D-value of the average charging current and temperature between the two charging data segments used in Equation (\ref{Eq:relative error-S1}) and (\ref{Eq:error cal-S2}) must be less than the predefined threshold.

\textcircled{5} EEM error calculation and uncertainty estimation: The EEM error of a FCS in comparison chain is calculated according to Equation (\ref{Eq:error cal-S2}). The uncertainty estimation method is detailed and introduced in Section \ref{Subsection: uncertainty}.

\subsection{Uncertainty analysis}\label{Subsection: uncertainty}

\begin{figure}[tp!]
    \centering
    \includegraphics[width=0.4\textwidth]{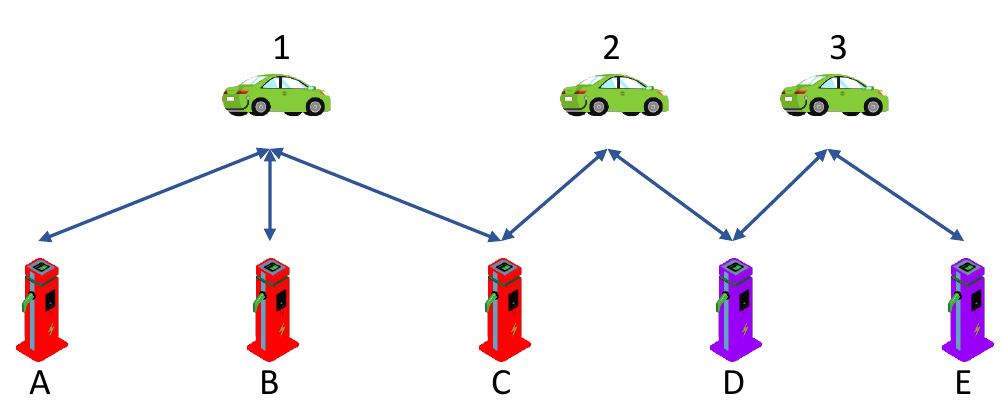}
    \caption{An example for MPC method. Vehicles and piles connected by blue arrows indicate the presence of charging data segments.}
    \label{fig: estimation example}
\end{figure}

In order to facilitate the subsequent discussion, as shown in Fig.~\ref{fig: estimation example}, this section illustrates the uncertainty components in the MPC method by using an example with 3 vehicles and 5-FCSs. For a pair of EV-FCS (connected by a blue arrow), there are two charging data segments. For EV-1 and FCS-A, it is assumed that the corresponding EEM values of the FCS-A for the two charging data segments are denoted as $E_{\mathrm{A1-1}}$ and $E_{\mathrm{A1-2}}$, respectively. While the changes of SOC measured by the BMS are denoted as $\mathrm{\Delta SOC_{A1-1}}$ and $\mathrm{\Delta SOC_{A1-2}}$, respectively. The recording method for the remaining charging data segments follows the same pattern. Furthermore, based on the charging process data, the range of the SOC quantization error is narrowed down to $\left[ {{y_{\min }},{y_{\max }}} \right]$. Referring to Equation (\ref{Eq: expected Ed-S3}), the expected values of BPED are as given by
\begin{align}\label{Eq: EdA1-1-S4}
    &{E_{\rm{dA1-1}}} =\eta _{\mathrm{A1-1}}{E^\prime }_{{\rm{dA}}1 - 1}=
    \notag
    \\&\eta _{\mathrm{A1-1}} \int_{{y_{\min }}}^{{y_{\max }}} {\frac{E_{\mathrm{A1-1}}}{{\Delta {\mathrm{SOC_{A1-1}}} + y}}} {f_{y|({y_{\min }} \le y \le {y_{\max }})}}{\rm{d}}y,
\end{align}
\begin{align}\label{Eq: EdA1-2-S4}
    &{E_{\rm{dA1-2}}} =\eta _{\mathrm{A1-2}}{E^\prime }_{{\rm{dA}}1 - 2}=
    \notag
    \\&\eta _{\mathrm{A1-2}} \int_{{y_{\min }}}^{{y_{\max }}} {\frac{E_{\mathrm{A1-2}}}{{\Delta {\mathrm{SOC_{A1-2}}} + y}}} {f_{y|({y_{\min }} \le y \le {y_{\max }})}}{\rm{d}}y.
\end{align}
$\eta _{\mathrm{A1-1}}$ and $\eta _{\mathrm{A1-1}}$ are conversion efficiency.  The uncertainty for $E_{\rm{dA1-1}}$ and $E_{\rm{dA1-2}}$ encompass the estimation error for conversion efficiency (UNC-1), SOC quantization error (UNC-2) and the repeatability error of BPED (UNC-3). Consequently, the assessment of the uncertainty components is developed.
\begin{enumerate}
    \item Estimation error for conversion efficiency: According to the investigation results, the resistance of the connection cable is approximately 2.2\,$\mathrm{m\Omega}$ (for a length of 5\,m). And the contact resistance at the cable connections is about 0.1\,$\mathrm{m\Omega}$. The charging voltage of FCSs can exceed 400\,V, and the charging current typically remains below 200\,A. Consequently, the conversion efficiency of FCSs is higher than 99.88\%. In light of this, within the MPC model, the conversion efficiency of charging stations is set to 1, with a relative estimation error of approximately 0.2\%.
    \item SOC quantization error: Referring to the SOC quantization error model established in Section \ref{Subsection: SOC quantization}, the Type-B uncertainty assessment method is employed, which utilizes the standard deviation of ${E^\prime }_{{\rm{dA}}1 - 1}$ and ${E^\prime }_{{\rm{dA}}1 - 2}$ to characterize the uncertainty component arising from the SOC quantization error. The standard deviation of $E_{\rm{dA1-1}}$ is given by
    \begin{equation}\label{Eq: sigma-S4}
        \sigma _1({E^\prime }_{{\rm{dA}}1 - 1}) = \sqrt {{\rm{E}}({E^\prime }_{\rm{dA1-1}}^2) - {{\rm{E}}^2}({{E^\prime }_{\rm{dA1-1}}})}.
    \end{equation}
    In which, ${\rm{E}}({E^\prime }_{\rm{dA1-1}}^2)$ is calculated by
    \begin{align}\label{Eq: Ed2-S4}
        &{\rm{E}}({E^\prime }_{\rm{dA1-1}}^2) = 
        \notag
        \\&{\int_{{y_{\min }}}^{{y_{\max }}} {\left( {\frac{E_{\mathrm{dA1-1}}}{{\Delta {\mathrm{SOC_{A1-1}}} + y}}} \right)} ^2}{f_{y|({y_{\min }} \le y \le {y_{\max }})}}{\rm{d}}y.
    \end{align}
    The details of the integration are shown in Table \ref{Table: Ed detail}. Thus, $\sigma _1({E^\prime }_{{\rm{dA}}1 - 1})$ can be calculated, so as $\sigma _1({E^\prime }_{{\rm{dA}}1 - 2})$
    \item Repeatability error of BPED: The TyAn uncertainty assessment method is employed. According to tests in Section \ref{Subsection: Repeatability error of BPED}, the relative standard deviation of BPED is below 6\%. However, the average values are used in Equation (\ref{Eq: EdA1-1-S4}) and (\ref{Eq: EdA1-2-S4}), hence, the uncertainty component of $E_{\rm{dA1-1}}$ and $E_{\rm{dA1-2}}$ arising from BPED repeatability error is given by
    \begin{equation}\label{sigma2-EdA1-1-S4}
        {\sigma _2}\left( {E^\prime }_{{\rm{dA}}1 - 1} \right) = \frac{{{\rm{std}}\left( {{\rm{BPED}}} \right)}{E^\prime }_{{\rm{dA}}1 - 1}}{{\sqrt {\Delta {\rm{SO}}{{\rm{C}}_{{\rm{A}}1 - 1}}} }},
    \end{equation}
    \begin{equation}\label{sigma2-EdA1-2-S4}
        {\sigma _2}\left( {E^\prime }_{{\rm{dA}}1 - 2} \right) = \frac{{{\rm{std}}\left( {{\rm{BPED}}} \right)}{E^\prime }_{{\rm{dA}}1 - 2}}{{\sqrt {\Delta {\rm{SO}}{{\rm{C}}_{{\rm{A}}1 - 2}}} }}.
    \end{equation}
    In which, ${{\rm{std}}\left( {{\rm{BPED}}} \right)}$ is the relative uncertainty of BPED.
\end{enumerate}

Here, the comprehensive uncertainties of ${E^\prime }_{{\rm{dA}}1 - 1}$ and ${E^\prime }_{{\rm{dA}}1 - 2}$ are given by
\begin{equation}\label{sigma-EdA1-1-S4}
    \sigma \left( {E^\prime }_{{\rm{dA}}1 - 1} \right) = \sqrt {\left( \sigma _1^2\left( {E^\prime }_{{\rm{dA}}1 - 1} \right) + \sigma _2^2\left( {E^\prime }_{{\rm{dA}}1 - 1} \right) \right)},
\end{equation}
\begin{equation}\label{sigma-EdA1-2-S4}
    \sigma \left( {E^\prime }_{{\rm{dA}}1 - 2} \right) = \sqrt {\left( \sigma _1^2\left( {E^\prime }_{{\rm{dA}}1 - 2} \right) + \sigma _2^2\left( {E^\prime }_{{\rm{dA}}1 - 2} \right) \right)}.
\end{equation}
Furthermore, the uncertainties of ${E}_{{\rm{dA}}1 - 1}$ and ${E}_{{\rm{dA}}1 - 2}$ are given by
\begin{equation}\label{sigma-EdA1-1-f-S4}
    \sigma \left( {{E_{{\rm{dA}}1 - 1}}} \right) = {E_{{\rm{dA}}1 - 1}}\sqrt {\sigma _{{\rm{r}} - {\rm{cv}}}^2 + \frac{{\sigma {{\left( {{E^\prime }_{{\rm{dA}}1 - 1}} \right)}^2}}}{{{{\left( {{E^\prime }_{{\rm{dA}}1 - 1}} \right)}^2}}}} ,
\end{equation}
\begin{equation}\label{sigma-EdA1-2-f-S4}
    \sigma \left( {{E_{{\rm{dA}}1 - 2}}} \right) = {E_{{\rm{dA}}1 - 2}}\sqrt {\sigma _{{\rm{r}} - {\rm{cv}}}^2 + \frac{{\sigma {{\left( {{E^\prime }_{{\rm{dA}}1 - 2}} \right)}^2}}}{{{{\left( {{E^\prime }_{{\rm{dA}}1 - 2}} \right)}^2}}}},
\end{equation}
where ${\sigma _{{\rm{r - cv}}}}$ the relative uncertainty of the onversion efficiency. The BPED of EV-1, determined by FCS-A, is given by
\begin{equation}\label{Eq: EdA1}
    {E_{\rm{dA1}}} ={\frac{E_{\rm{dA1-1}}+E_{\rm{dA1-2}}}{2}}.
\end{equation}
Therefore, the uncertainty for $E_{\rm{dA1}}$ is given by
\begin{equation}\label{Eq: sig_EdA1}
    \sigma \left( {{E_{{\rm{dA1}}}}} \right) = \sqrt {\frac{{\sigma {{\left( {{E_{{\rm{dA1}}- 1}}} \right)}^2} + \sigma {{\left( {{E_{{\rm{dA1}}- 2}}} \right)}^2}}}{4}}.
\end{equation}
Herein, employing the same computational methodology, the value of $\sigma \left( {{E_{{\rm{dA2}}}}} \right)$, $\sigma \left( {{E_{{\rm{dB1}}}}} \right)$, and etc. can be calculated.

Subsequently, the RCSs need to be identified. As shown in Fig.~\ref{fig: estimation example}, EV-1 has charging records at FCS-A, FCS-B, and FCS-C. Consequently, by utilizing the charging data, it is feasible to estimate the true value of BPED of EV-1. Assuming that the relative EEM errors of the three FCSs do not exceed $l$, and that their EEM errors are distributed within the interval of $\left[ { - \frac{l}{2},\frac{l}{2}} \right]$, the estimated value of BPED is given by 
\begin{equation}\label{Eq: Ed1-S4}
    {E_{{\rm{d}}1}} = \frac{{{E_{{\rm{dA}}1}} + {E_{{\rm{dB}}1}} + {E_{{\rm{dC}}1}}}}{3}.
\end{equation}
Therefore, the uncertainty of ${E_{{\rm{d}}1}}$ is given by
\begin{equation}\label{Eq: sig_Ed1-S4}
    \sigma \left( {{E_{{\rm{d}}1}}} \right) = \frac{{\sqrt {\sigma {{\left( {{E_{{\rm{dA}}1}}} \right)}^2} + \sigma {{\left( {{E_{{\rm{dB}}1}}} \right)}^2} + \sigma {{\left( {{E_{{\rm{dC}}1}}} \right)}^2}} }}{3}.
\end{equation}
In addition, due to the normal distribution of the errors of FCSs, the estimation error of BPED  (UNC-3) should not be neglected. According to Equation (\ref{Eq: RCS-error}), the estimation error of $E_{\mathrm{d1}}$ is
\begin{equation}\label{Eq: ep-RCS-error-S4}
    e = \frac{{{\gamma _{\rm{A}}} + {\gamma _{\rm{B}}} + {\gamma _{\rm{C}}}}}{3}{E_{{\rm{d}}10}}.
\end{equation}
In which, ${E_{{\rm{d}}10}}$ is the true value of EV-1's BPED. According to Fig. \ref{fig: distribution of error}, the EEM of a FCS complies the normal distribution as indicated in
\begin{equation}\label{Eq: F(x)-S4}
    {f_{{\upgamma}}}\left( x \right) = \frac{1}{{\sqrt {2{\mathrm{\uppi }}} \sigma }}{{\rm{e}}^{ - \frac{{{{ {x} }^2}}}{{2{\sigma ^2}}}}}.
\end{equation}
Under the aforementioned conditions, the probability density of $\gamma _{\mathrm{A}}$, $\gamma _{\mathrm{B}}$ and $\gamma _{\mathrm{C}}$ is given by
\begin{equation}\label{Eq: Fl(x)-S4}
    {f_{\upgamma |l}}\left( x \right) = \left\{ {\begin{array}{*{20}{l}}
{\frac{{{f_\upgamma }\left( x \right)}}{{\int_{ - \frac{l}{2}}^{\frac{l}{2}} {{f_\upgamma }\left( x \right){\rm{d}}x} }}}&{, - \frac{l}{2} \le x \le \frac{l}{2}}\\
0&{,{\rm{others}}}
\end{array}} \right..
\end{equation}
Therefore, the uncertainty component arising from the estimation error of ${{E_{{\rm{d10}}}}}$ is given by 
\begin{equation}\label{Eq: ep-RCS-error-S4}
    \sigma \left( e \right) = \frac{{\sigma \sqrt {{\rm{erf}}\left( {\frac{l}{{2\sqrt 2 \sigma }}} \right)} }}{{\sqrt {3d} }}{E_{{\rm{d}}10}}.
\end{equation}
Furthermore, the EEM error of the RCS-A is given by
\begin{equation}\label{Eq: rA-ori-S4}
    {\gamma _{\rm{A}}} = \frac{{{E_{{\rm{dA}}1}} - {E_{{\rm{d}}10}}}}{{{E_{{\rm{d}}10}}}}= \frac{{{E_{{\rm{dA}}1}}}}{{{E_{{\rm{d}}1}} - e}} - 1.
\end{equation}

Apparently, the contributors to the error of $\gamma_{\mathrm{A}}$ encompass the errors of ${E_{{\rm{dA}}1}}$, ${E_{{\rm{d}}1}}$ and $e$. Owing to the non-linear relationship of these variables, it is challenging to directly employ Equation (\ref{Eq: rA-ori-S4}) to establish the uncertainty relationship between $\gamma_{\mathrm{A}}$ and ${E_{{\rm{dA}}1}}$, ${E_{{\rm{d}}1}}$, $e$. Herein, Equation (\ref{Eq: rA-ori-S4}) is rewritten using Taylor series expansion at $\left( {{E_{{\rm{dA}}1}^0},{E_{{\rm{d}}1}^0},{e^0}} \right)$, i.e.
{\small
\begin{align}\label{Eq: rA-S4}
    {\gamma _{\rm{A}}} &= \gamma _{\rm{A}}^0 + \frac{1}{{E_{{\rm{d}}1}^0 - {e^0}}}\left( {{E_{{\rm{dA}}1}} - E_{{\rm{dA}}1}^0} \right)
    \notag
    \\& - \frac{{E_{{\rm{dA}}1}^0}}{{{{\left( {E_{{\rm{d}}1}^0 - {e^0}} \right)}^2}}}\left( {{E_{{\rm{d}}1}} - E_{{\rm{d}}1}^0} \right) + \frac{{E_{{\rm{dA}}1}^0}}{{{{\left( {E_{{\rm{d}}1}^0 - {e^0}} \right)}^2}}}\left( {e - {e^0}} \right). 
\end{align}
}
${E_{{\rm{dA}}1}^0}$ and ${E_{{\rm{d}}1}^0}$ are calculated by Equation (\ref{Eq: EdA1-1-S4}) and (\ref{Eq: Ed1-S4}), $e^0$ is assigned as 0. $\gamma _{\rm{A}}^0$ is calculated by Equation (\ref{Eq: rA-ori-S4}). Thus, the uncertainty of ${\gamma _{\rm{A}}}$ can be expressed as
{\small
\begin{align}\label{Eq: sig-rA-S4}
    \sigma \left( {{\gamma _{\rm{A}}}} \right)& = {\rm{sqrt}}\left( {\frac{{\sigma {{\left( {{E_{{\rm{dA}}1}}} \right)}^2}}}{{{{\left( {E_{{\rm{d}}1}^0 - {e^0}} \right)}^2}}} + \frac{{{{\left( {E_{{\rm{dA}}1}^0} \right)}^2}\sigma {{\left( {{E_{{\rm{d}}1}}} \right)}^2}}}{{{{\left( {E_{{\rm{d}}1}^0 - {e^0}} \right)}^4}}} + }\right.
    \notag
    \\&\left. {\frac{{{{\left( {E_{{\rm{dA}}1}^0} \right)}^2}\sigma {{\left( e \right)}^2}}}{{{{\left( {E_{{\rm{d}}1}^0 - {e^0}} \right)}^4}}} - \frac{{2E_{{\rm{dA}}1}^0}}{{{{\left( {E_{{\rm{d}}1}^0 - {e^0}} \right)}^3}}}{\rm{COV}}\left( {{E_{{\rm{dA}}1}},{E_{{\rm{d}}1}}} \right)} \right). 
\end{align}}

Due to the lack of independence between ${E_{{\rm{dA}}1}}$ and ${E_{{\rm{d}}1}}$, it is necessary to take into account the covariance in Equation (\ref{Eq: sig-rA-S4}). Here,
{\small
\begin{align}\label{Eq: COV-S4}
    {\rm{COV}}&\left( {{E_{{\rm{dA}}1}},{E_{{\rm{d}}1}}} \right) 
    \notag
    \\&= {\rm{E}}\left[ {\left( {{E_{{\rm{dA}}1}} - {\rm{E}}\left( {{E_{{\rm{d}}1}}} \right)} \right)\left( {{E_{{\rm{d}}1}} - {\rm{E}}\left( {{E_{{\rm{dA}}1}}} \right)} \right)} \right]
    \notag
    \\& = {\rm{E}}\left( {{E_{{\rm{dA}}1}}{E_{{\rm{d}}1}}} \right) - {\rm{E}}\left( {{E_{{\rm{d}}1}}} \right){\rm{E}}\left( {{E_{{\rm{dA}}1}}} \right). 
\end{align}}
Substitute Equation (\ref{Eq: Ed1-S4}) into Equation (\ref{Eq: COV-S4}).In which, 
\begin{align}\label{Eq: COV-E-S4}
    {\rm{E}}\left( {{E_{{\rm{dA}}1}}{E_{{\rm{d}}1}}} \right)& =
    \notag
    \\&\frac{1}{3}{\rm{E}}\left( {E_{{\rm{dA}}1}^2 + {E_{{\rm{dA}}1}}{E_{{\rm{dB}}1}} + {E_{{\rm{dA}}1}}{E_{{\rm{dC}}1}}} \right).
\end{align}
Generally speaking, the charging behaviors among different FCSs are mutually independent. Therefore, Equation (\ref{Eq: COV-E-S4}) can be expanded as
\begin{align}\label{Eq: COV-E-expa-S4}
    {\rm{E}}&\left( {{E_{{\rm{dA}}1}}{E_{{\rm{d}}1}}} \right) = \frac{1}{3}\left[ {{\rm{E}}\left( {E_{{\rm{dA}}1}^2} \right) + }\right.
    \notag
    \\&\left. {{\rm{E}}\left( {{E_{{\rm{dA}}1}}} \right){\rm{E}}\left( {{E_{{\rm{dB}}1}}} \right) + {\rm{E}}\left( {{E_{{\rm{dA}}1}}} \right){\rm{E}}\left( {{E_{{\rm{dC}}1}}} \right)} \right].
\end{align}
In which,
\begin{equation}\label{Eq: E(ED2)-S4}
    {\rm{E}}\left( {E_{{\rm{dA}}1}^2} \right) = \sigma {\left( {{E_{{\rm{dA}}1}}} \right)^2} + {\rm{E}}{\left( {{E_{{\rm{dA}}1}}} \right)^2}.
\end{equation}
Up to now, $\gamma _{\rm{A}}$ and $\sigma \left( {{\gamma _{\rm{A}}}} \right)$ can both be calculated. So as ${\gamma _{\mathrm{B}}}$, $\sigma \left( {{\gamma _{\rm{B}}}} \right)$ and ${\gamma _{\mathrm{C}}}$, $\sigma \left( {{\gamma _{\rm{C}}}} \right)$. Then, by employing the comparison chains, the EEM errors of other FCSs can be calculated. In this case, by EV-2 and EV-3, a comparison chain FCS-C$\leftrightarrow$FCS-D$\leftrightarrow$FCS-E is established. Thus, the EEM errors for FCS-D and FCS-E are given by
\begin{equation}\label{Eq: rD-ori-S4}
    {\gamma _{\rm{D}}} = \frac{{{E_{{\rm{dD2}}}} - {E_{{\rm{dC2}}}}}}{{{E_{{\rm{dC2}}}}}} + \frac{{{E_{{\rm{dD2}}}}}}{{{E_{{\rm{dC2}}}}}}{\gamma _{\rm{C}}},
\end{equation}
\begin{equation}\label{Eq: rE-ori-S4}
    {\gamma _{\rm{E}}} = \frac{{{E_{{\rm{dE3}}}} - {E_{{\rm{dD3}}}}}}{{{E_{{\rm{dD3}}}}}} + \frac{{{E_{{\rm{dE3}}}}}}{{{E_{{\rm{dD3}}}}}}{\gamma _{\rm{D}}}.
\end{equation}

Similar to the computation process of $\sigma \left( {{\gamma _{\rm{A}}}} \right)$, the issue of nonlinear relationships arises in the computation processes of $\sigma \left( {{\gamma _{\rm{D}}}} \right)$ and $\sigma \left( {{\gamma _{\rm{E}}}} \right)$. By emulating Equation (\ref{Eq: rA-S4}), Equation (\ref{Eq: rD-ori-S4}) is expanded in a Taylor series at $\left( {E_{{\rm{dD}}2}^0,E_{{\rm{dC}}2}^0,\gamma _{\rm{C}}^0} \right)$,
{\small
\begin{align}\label{Eq: rD-S4}
    {\gamma _{\rm{D}}} &= \gamma _{\rm{D}}^0 + \frac{{1 + \gamma _{\rm{C}}^0}}{{E_{{\rm{dC}}2}^0}}\left( {{E_{{\rm{dD}}2}} - E_{{\rm{dD}}2}^0} \right)+
    \notag
    \\& \frac{{\left( {1 + \gamma _{\rm{C}}^0} \right)E_{{\rm{dD}}2}^0}}{{{{\left( {E_{{\rm{dC}}2}^0} \right)}^2}}}\left( {{E_{{\rm{dC}}2}} - E_{{\rm{dC}}2}^0} \right) + \frac{{E_{{\rm{dD}}2}^0}}{{E_{{\rm{dC}}2}^0}}\left( {{\gamma _{\rm{C}}} - \gamma _{\rm{C}}^0} \right). 
\end{align}}
Thus, the uncertainty of ${\gamma _{\rm{D}}}$ is given by
{\small
\begin{align}\label{Eq: sig_rD-S4}
     &\sigma \left( {{\gamma _{\rm{D}}}} \right) = {\rm{sqrt}}\left( {{{\left( {\frac{{1 + \gamma _{\rm{C}}^0}}{{E_{{\rm{dC}}2}^0}}} \right)}^2}\sigma {{\left( {{E_{{\rm{dD}}2}}} \right)}^2}} \right. 
    \notag
    \\& \left. { + {{\left( {\frac{{\left( {1 + \gamma _{\rm{C}}^0} \right)E_{{\rm{dD}}2}^0}}{{{{\left( {E_{{\rm{dC}}2}^0} \right)}^2}}}} \right)}^2}\sigma {{\left( {{E_{{\rm{dC}}2}}} \right)}^2} + {{\left( {\frac{{E_{{\rm{dD}}2}^0}}{{E_{{\rm{dC}}2}^0}}} \right)}^2}\sigma {{\left( {{\gamma _{\rm{C}}}} \right)}^2}} \right).
\end{align}}

The same methodology can be employed to calculate $\sigma \left( {{\gamma _{\rm{E}}}} \right)$. At this juncture, the estimation of EEM errors for FCSs as well as the uncertainty associated with the estimation results are completed. Furthermore, it is imperative to discriminate whether there are EEM performance defects at FCSs. It is postulated that if the absolute value of EEM error is less than $\gamma _{\mathrm{t}}$, the EEM error is deemed acceptable. Conversely, if the EEM error exceeds this threshold, it is considered unacceptable. Thus, the acceptable EEM error range is defined as $\left[ { - {\gamma _{\rm{t}}},{\gamma _{\rm{t}}}} \right]$. The MPC method is capable of delineating the distribution interval of the EEM error, namely $\left[ {{\gamma _{{\rm{FCS}}}} - \sigma \left( {{\gamma _{{\rm{FCS}}}}} \right),{\gamma _{{\rm{FCS}}}} + \sigma \left( {{\gamma _{{\rm{FCS}}}}} \right)} \right]$. Accordingly, this paper defines the probability of acceptable EEM error, which is defined as the ratio of the length of the EEM error distribution interval that falls within the acceptable EEM error range to the total length of the distribution interval. For instance, if ${\gamma _{{\rm{FCS}}}} - \sigma \left( {{\gamma _{{\rm{FCS}}}}} \right) <  - {\gamma _{\rm{t}}}$ and $ - {\gamma _{\rm{t}}} < {\gamma _{{\rm{FCS}}}} + \sigma \left( {{\gamma _{{\rm{FCS}}}}} \right) < {\gamma _{\rm{t}}}$, the probability of acceptable EEM error is
\begin{equation}\label{Eq: pro-accept-S4}
    {\rm{P}} = \frac{{{\gamma _{{\rm{FCS}}}} + \sigma \left( {{\gamma _{{\rm{FCS}}}}} \right) + {\gamma _{\rm{t}}}}}{{2\sigma \left( {{\gamma _{{\rm{FCS}}}}} \right)}} \times 100\% .
\end{equation}
When this probability value exceeds 50\%, the EEM error is deemed acceptable. Especially, when the EEM error distribution interval encompasses the acceptable range, that is, ${\gamma _{{\rm{FCS}}}} - \sigma \left( {{\gamma _{{\rm{FCS}}}}} \right) <  - {\gamma _{\rm{t}}}$ and ${\gamma _{{\rm{FCS}}}} + \sigma \left( {{\gamma _{{\rm{FCS}}}}} \right) > {\gamma _{\rm{t}}}$, the estimation result is deemed unreliable.

\section{Estimation model validation}\label{Section: validation}
Based on the principles elucidated in the preceding sections, this section employs field operational data to undertake the validation of the effectiveness of the MPC method. The charging data utilized in this section are derived from a subset of operational FCSs (with accuracy grade of 2\%) located in North China. The timestamps of all charging data fall within March 2024. The dataset encompasses 7195 items of charging data, 567 FCSs, and 1274 EVs. The parameter settings in the estimated model are represented in Table \ref{Table-parameters setting}.
\begin{table*}[tp!]
\caption{Parameters of the estimation model}
\begin{center}
\begin{tabular}{c|c|c|c}
\hline
Parameter & Value & Parameter & Value\\
\hline
Charging current threshold & 4~A & Minimum $\Delta {\mathrm{SOC}}$ & 20\%\\
\hline
Relative repeatability error of BPED & 6\% & Minimum number of FCSs (for RCSs) & 3\\
\hline
Relative repeatability error threshold of expected BPED & 1\% & $\sigma$ for EEM error distribution of FCS & 1.62\%\\
\hline
Relative EEM error threshold (for RCSs) & 0.67\% & Longest comparison chain & 4 FCSs\\
\hline
D-temperature threshold & 5~\textcelsius & Acceptable EEM error & $\left[ { - 2\%, 2\% } \right]$\\
\hline
\end{tabular}
\label{Table-parameters setting}
\end{center}
\end{table*}

\begin{figure*}[tp!]
    \centering
    \includegraphics[width=0.75\textwidth]{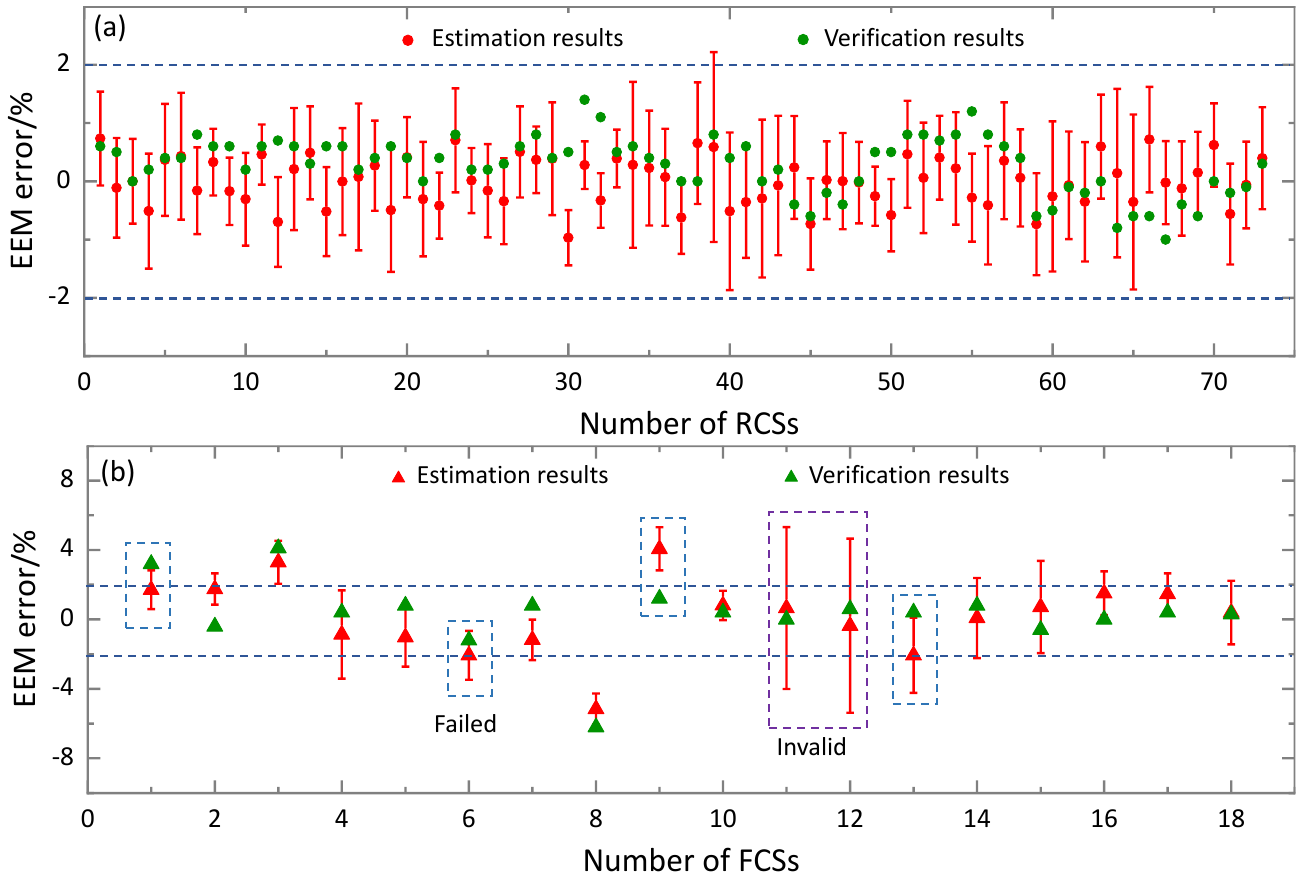}
    \caption{A comparison between the estimation results and the verification results is presented, wherein (a) depicts the comparison for the RCSs, with 73 RCSs possessing verification results, and (b) illustrates the comparison for the other FCSs, with 18 FCSs having verification results.}
    \label{fig: estimation results}
\end{figure*}

According to the MPC method, the number of RCSs is 423, among them, the number of RCSs possessed verification results is 73. The number of comparison chains is 294. The number of other FCSs with estimation results is 144, with 18 of them having verification results. The comparison between the estimation and verification results is shown in Fig.~\ref{fig: estimation results}. Wherein, the EEM errors of all RCSs are estimated as acceptable, aligning perfectly with the verification results. For other FCSs, the estimation EEM errors for FCS-11 and FCS-12 exhibited large uncertainty, rendering the results invalid. The estimation EEM error for FCS-1 is deemed acceptable (with an expected EEM error of -1.7\%, an uncertainty of 1.1\%, and an acceptable probability of 63.1\%), whereas the verification result is unacceptable (with a verification EEM error of 3.2\%). The EEM errors of FCS-6, FCS-9, and FCS-13 are estimated as unacceptable (with expected EEM errors of -2.1\%, -4.1\%, and -2.1\%, respectively, uncertainties of 1.4\%, 1.2\%, and 2.1\%, and acceptable-probabilities of 47.6\%, 0, and 48.3\%, respectively), whereas their verification results are acceptable (with verification EEM errors of -1.2\%, 1.2\%, and 0.4\%, respectively). The estimation conclusions for the remaining FCSs correspond with their verification conclusions.

\section{Conclusion}\label{Section: conclusion}
To overcome the drawbacks of high costs and low efficiency in existing on-site verification methods, this paper proposes the MPC based on the concept of digital metrology. In MPC method, the comparison chains for EEM performance among multiple FCSs are established and mediated by SOC, enabling EEM error estimation of large-scale FCSs. Additionally, considering actual operational conditions, this paper summarizes the interfering factors of estimation results, including charge-discharge cycles, temperature, charging current, SOC estimation accuracy, and SOC quantization errors. The corresponding error models and uncertainty models are established. Also, a method for discriminating whether there are EEM performance defects is proposed. Finally, the feasibility of MPC method is validated, with results indicating that for FCSs with an accuracy grade of 2\%, the discriminative accuracy exceeds 95\%. 

Looking ahead, with technological innovations in SOC estimation algorithms, there is potential for further enhancing the accuracy of SOC estimation. Moreover, the standardization of charging data in EV industry continues to strengthen. Hence, the accuracy of EEM error estimation can be improved further, facilitating more accurate online monitoring of large-scale FCSs for EEM performance. Ultimately, the MPC method is expected to promote the digital transformation of FCSs measurement verification.

\end{document}